\documentclass[twocolumn]{aa}
\usepackage{epsfig}
\usepackage{graphicx}
\usepackage[latin1]{inputenc}
\def\simlt{\ \raise -2.truept\hbox{\rlap{\hbox{$\sim$}}\raise5.truept   %
\hbox{$<$}\ }}
\def\simgt{\ \raise -2.truept\hbox{\rlap{\hbox{$\sim$}}\raise5.truept   %
\hbox{$>$}\ }}                                                          %

\def\be{\begin{equation}}
\def\ee{\end{equation}}
\def\newline{\hfil\break}

\def\la{\mathrel{\hbox{\rlap{\hbox{\lower4pt\hbox{$\sim$}}}\hbox{$<$}}}}
\def\ga{\mathrel{\hbox{\rlap{\hbox{\lower4pt\hbox{$\sim$}}}\hbox{$>$}}}}

\begin{document}
\title{Probing photon decay with the Sunyaev-Zel'dovich effect}
   \author{S. Colafrancesco\inst{1,2} and P. Marchegiani\inst{1,2}}
   \offprints{S. Colafrancesco}
 \institute{INAF - Osservatorio Astronomico di Roma
              via Frascati 33, I-00040 Monteporzio, Italy.
              Email: sergio.colafrancesco@oa-roma.inaf.it
              \and
              School of Physics, University of the Witwatersrand, Private Bag 3, 2050-Johannesburg, South Africa.
              Email: sergio.colafrancesco@wits.ac.za
             }
\date{Received / Accepted }
\authorrunning {S. Colafrancesco et al.}
\titlerunning {SZ effect and photon decay}
\abstract
   {The fundamental properties of the photon have deep impact on the astrophysical processes that involve it, like the inverse Compton scattering of CMB photon by energetic electrons residing within galaxy cluster atmospheres, usually referred to as the Sunyaev-Zel'dovich effect (SZE).}
   {We calculate the combined constraints on the photon decay time and mass by studying the impact  
of the modified CMB spectrum, as recently calculated (Heeck 2013), on the SZE of galaxy clusters.}
  {We analyze the modifications of the SZE as produced by photon decay effects. 
We study in details the frequency regimes where these modifications are large and where the constraints derived from the SZE can be stronger with respect to those already obtained from the CMB spectrum.}
   {We show that the SZE can set limits on the photon decay time and mass, or on $E^* = \frac{t_0}{\tau_\gamma}m_\gamma c^2$, that are stronger than those obtained from the CMB: the main constraints come from the low frequency range $\nu \approx 10-50$ GHz where  the modified SZE $\Delta I_{mod}$ is larger than the standard one $\Delta I$, with the difference $|(\Delta I_{mod} - \Delta I)|$ increasing with the frequency for increasing values of $E^*$; additional constraints can be set in the range $120 - 180$ GHz where there is an increase of the frequency position of the minimum of $\Delta I_{mod}$ with respect to the standard one with increasing values of $E^*$.}
  {We demonstrated that the effect of photon decay can be measured or constrained by the Square Kilometer Array in the optimal range $\approx 10-30$ GHz setting limits of $E^* \simlt 1.4 \times 10^{-9}$ eV and $5 \times 10^{-10}$ eV for 30 and 260 hour integration for A2163, respectively. These limits are stronger than those obtained with the COBE-FIRAS spectral measurements of the CMB.}

 \maketitle

\section{Introduction}
 \label{sec.intro}

The possibility of having a non-zero mass for the photon is a valid theoretical possibility described by the Proca (1936) equations that extend Maxwell equations to the case of a massive photon. Even in the context of quantum theories of electro-magnetic fields, the gauge invariance break caused by a non-zero photon mass can be repaired via the  St\"uckelberg (1957) mechanism.\\ 
%
Experimental limits on the photon mass can be derived from laboratory experiments on the Coulomb law and on magnetic fields (see discussion in Heeck 2013), with more stringent limits that can be derived by using magnetic fields on extra terrestrial scales. 
In fact, the most stringent limit available today, 
$m_\gamma c^2 < 10^{-18} \mbox{ eV}$, is derived from observations of the magnetic field in the solar wind (Ryutov 2007). Observations of magnetic fields on larger scales can yield even stronger (although still uncertain) limits (see Goldhaber \& Nieto 2010).\\
More recently Heeck (2013) has discussed also the possibility that a photon might decay in lighter particles  (e.g., the lightest neutrino or even particles outside the standard model), showing that - contrary to the limits on the photon mass - there are not yet stringent experimental limits on the photon decay time. In his work, Heeck (2013) derived a first limit on the photon decay time comparing the distortion on the Cosmic Microwave Background (CMB) spectrum induced by the photon decay with the COBE-FIRAS data. 
The constraints derived from the CMB yield rather weak limits on the photon mass, but very competitive limits on the photon decay time that are stronger than the existing ones: 
\begin{equation}
\tau_\gamma > 2\times10^{-10} \left( \frac{m_\gamma c^2}{10^{-18} \rm{ eV}} \right) t_0
\end{equation}
where $t_0=13.8\times10^9$ yr is the age of the universe.

In this work we want to explore the possibility of setting even stronger limits on the basic properties of the photon (i.e. its mass and decay time) by using the Sunyaev-Zel'dovich effect (hereafter SZE; Sunyaev \& Zel'dovich 1972, see Colafrancesco et al. 2003 for a generalized derivation), i.e. the spectral distortions produced by the inverse Compton scattering of CMB photons off high energy electrons (both thermal and non-thermal) that are present in clusters of galaxies and other cosmic structures (see Birkinshaw 1999, Colafrancesco 2012 for reviews).
Even though the SZE in clusters yields a distortion of the CMB spectrum typically of order $\sim 10^{-5}$, 
its ability to derive limits on the photon properties can be favored by the fact that it is a differential measure of spectral CMB distortions and therefore it can be easier to detect  a trace of the photon decay through specific SZE spectral measurements, or set more stringent limits on the photon properties than those obtained from the CMB spectrum analysis.
%

\section{The SZE with decaying CMB photons}


The effect of photon decay on the CMB spectrum can be calculated (see, e.g., Heeck 2013) and the final energy density of the photon population is given by 
\begin{eqnarray}
\rho(E,T_0)dE & \simeq & \frac{8\pi}{(hc)^3} \frac{E^3 dE}{\exp\left(\sqrt{E^2-m_\gamma^2c^4}/k_B T_0\right)-1} \times
 \nonumber\\
 & & \sqrt{1-\frac{m_\gamma^2c^4}{E^2}} \exp\left(-\frac{\Gamma_0}{\hbar c} \frac{m_\gamma c^2}{E} d_L \right)
 \label{cmb_density_mod}
\end{eqnarray}  
where $T_0$ is the CMB temperature,  $\Gamma_0=\hbar /\tau_\gamma$ is the decay width of the photon, $d_L=47\times10^9$ ly is the comoving distance of the last scattering surface.
In the following we define the quantity
\begin{equation}
E^*\equiv \frac{\Gamma_0}{\hbar} t_0 m_\gamma c^2=\frac{t_0}{\tau_\gamma}m_\gamma c^2
\end{equation}
that parametrizes the effect of the photon decay.
The limits on both the photon mass and its decay time  obtained by Heeck (2013), once combined together, provide a  limit on the quantity $E^*$ of
\begin{equation}
E^* < 5 \times 10^{-9} \mbox{ eV}.
\label{lim_estar}
\end{equation}
%
%
%
%
The spectral distortion of the CMB spectrum due to the SZE  is given by
\begin{equation}
I(x)=\int_{-\infty}^{+\infty} I_0(xe^{-s}) P(s) ds
\label{spettro_risultante}
\end{equation}
(Wright 1979, Colafrancesco et al. 2003), where $x=h\nu/(k_B T_0)$ is the normalized photon frequency, $P(s)$ is the photon redistribution function (yielding the probability of a logarithmic shift  $s=\ln (\nu'/\nu)$ in the photon frequency) that depends on the electron spectrum producing the CMB Comptonization, and $I_0(x)$ is the specific intensity of the incident CMB radiation field. 
%
The redistribution function $P(s)$ can be calculated at the desired approximation order in the plasma depth 
$\tau=\sigma_T \int n_e d\ell$ (where $n_e$ is the plasma electron density) or via a general relativistic method by using Fourier transform properties (see Colafrancesco et al. 2003 for details), that contains the relativistic corrections required for high temperature or relativistic electrons
(see, e.g., Wright 1979, Rephaeli 1995, Challinor \& Lasenby 1998, Itoh et al. 1998, Sazonov \& Sunyaev 1998), the effect of multiple scattering (see, e.g., Molnar \& Birkinshaw 1999, Dolgov et al. 2001, Itoh et al. 2001), and the treatment of multiple electron population combination.
In the following we calculate the SZE at second order in $\tau$ for the sake of having reasonably short computational times: we have verified, however, that our results  do not change appreciably even in the general, exact computation case. 
In the following we study the specific case of the SZE produced by a thermal electron population that provide the dominant contribution to the SZE observed in galaxy clusters with respect to the kinematic and non thermal SZEs (see, e.g., En\ss lin \& Kaiser 2000, Colafrancesco et al. 2003). 
We stress here that our general method also allows to take into account the additional contributions to the total SZE as produced by the kinematic SZE, by additional non-thermal electron populations, as well as other contributions to the SZE like line-of-sight variations of electron temperature and density (see, e.g., Colafrancesco \& Marchegiani 2010).
We also stress here that Chluba et al. (2013) and Chluba \& Dai (2013) showed that the effect of scattering anisotropies of the radiation field needs to be included in order to obtain correct results already at second order in $\tau$. We discuss in next Section the impact of the scattering anisotropies of the radiation field on our results.

The spectrum of the incident CMB radiation field $I_0(x)$ is derived from eq.(\ref{cmb_density_mod}) that describes the CMB energy density modified by the photon decay process
\begin{eqnarray}
I_{0,mod}(x)&=&2\frac{(k_B T_0)^3}{(hc)^2} \frac{x^3}{\exp \left[\sqrt{E^2-m_\gamma^2c^4}/E_0\right]-1}
\times \nonumber \\
 & & \sqrt{1-\frac{m_\gamma^2c^4}{E^2}} \exp\left[-\frac{E^*}{E}\frac{d_L}{ct_0}\right]
 \label{spettro_cmb_mod}
\end{eqnarray}
where $E_0=k_B T_0$ and $E=h\nu=xE_0$.
We note that for $m_\gamma=0$, which yields the value $E^*=0$, one obtains the original CMB spectrum without the photon decay effect: 
\begin{equation}
I_0(x)=2\frac{(k_B T_0)^3}{(hc)^2} \frac{x^3}{e^x-1}.
\end{equation}
Inserting the incoming modified CMB spectrum of eq. (\ref{spettro_cmb_mod}) in eq. (\ref{spettro_risultante}) we obtain the SZE spectrum modified by the photon decay process as
\begin{equation}
\Delta I_{mod}(x) = I_{mod}(x) - I_{0,mod}(x)
\label{SZE_modified}
\end{equation}
that can be calculate as a function of the relevant parameters  $m_\gamma$ and $\tau_\gamma$.
In the following we will use, for simplicity, the single parameter $E^*$ to quantify the changes in the SZE and in the CMB spectrum due to the photon decay. 
We note that in the non-relativistic limit and for low values of the electron temperature $T$ and optical depth $\tau$, the SZE spectrum in eq.(\ref{SZE_modified}) can be expressed in the form $\Delta I_{mod}(x)=2[(k_BT_0)^3/(hc)^2] y_0 g_{mod}(x)$, where $y_0=\sigma_T\int (k_B T/m_ec^2)n_e  d\ell$, and the function $g_{mod}(x)$ is given at the first order approximation in the parameter $\epsilon \equiv m_\gamma c^2/E_0\ll 1$ as:
\begin{eqnarray}
g_{mod}(x)&=&g(x)+ \epsilon \frac{d_L}{c \tau_\gamma} \left[ \frac{x^3}{e^x-1} \right] \left[ \frac{e^x(2+x)}{e^x-1} + \right. \nonumber  \\
& &\left. - \frac{2xe^{2x}}{(e^x-1)^2} +\frac{2}{x}\right],
\end{eqnarray}
where $g(x)$ is the non-relativistic SZE spectral function (see Zel'dovich \& Sunyaev 1969).

\begin{figure}[ht]
\begin{center}
{
 \epsfig{file=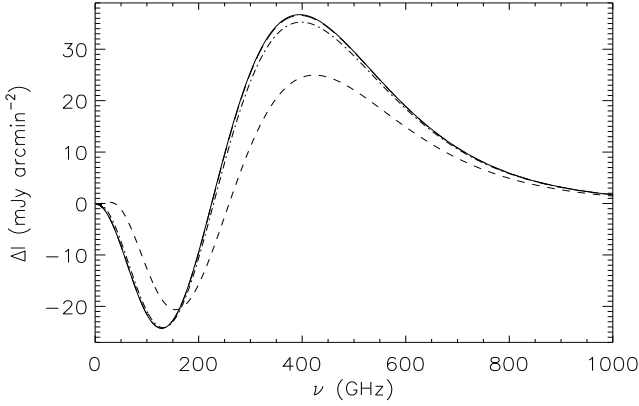,height=6.cm,width=9.cm,angle=0.0}
}
\end{center}
 \caption{\footnotesize{The thermal SZE (see eq.\ref{SZE_modified}) for $k_BT=15$ keV and $\tau=1\times10^{-2}$
 is shown for the case $E^*=0$ (solid curve, corresponding to the standard SZE), $10^{-6}$ eV (long dashed curve), $10^{-5}$ eV (dot-dashed curve) and $10^{-4}$ eV (dashed curve).
 }}
 \label{szmod_15}
\end{figure}

Fig. \ref{szmod_15} shows the spectrum of the thermal SZE for an electron population in a cluster with a temperature $k_B T=15$ keV and depth $\tau=1\times10^{-2}$ for various values of $E^*$.

\begin{figure}[ht]
\begin{center}
{
 \epsfig{file=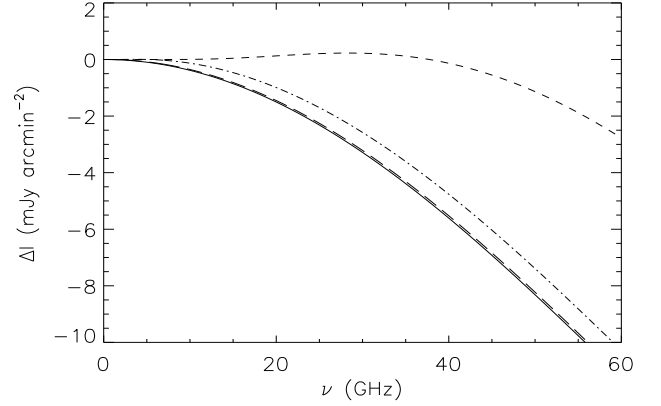,height=6.cm,width=9.cm,angle=0.0}
 \epsfig{file=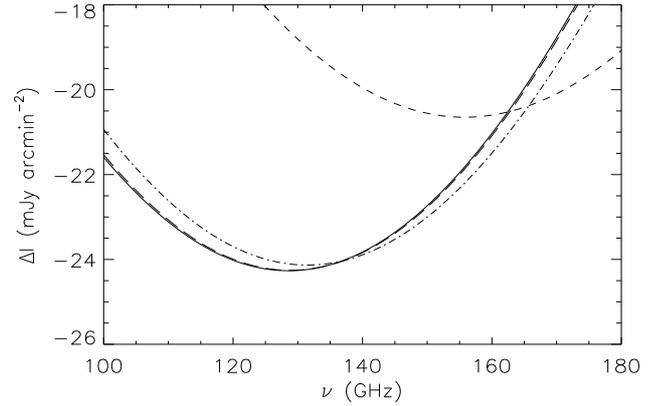,height=6.cm,width=9.cm,angle=0.0}
}
\end{center}
 \caption{\footnotesize{Zoom of the plot in Fig. \ref{szmod_15} for the two most interesting spectral regions (see text for details).
 }}
 \label{szmod_15_zoom}
\end{figure}

Fig. \ref{szmod_15_zoom} shows a zoom of the curves in Fig. \ref{szmod_15} in some interesting spectral regions: 
the low-$\nu$ region 
(0--60 GHz), and the frequency region around the minimum of the SZE in the range 100--180 GHz.
These spectral regions are interesting because there is a sensitive difference between the standard SZE  and that modified by the photon decay that becomes more evident with increasing values of $E^*$. 
It is clear, however, that the limit set by eq. (\ref{lim_estar}) indicates that the SZE  distortion must be small w.r.t. the standard amplitude of the SZE, and hence it also indicates that we need a more detailed theoretical analysis and a wider exploration of experimental techniques in order to determine the effect of photon decay on the SZE.
In the next Section we will discuss more specifically the possibilities of detecting the variations of the thermal SZE  due to the photon decay for various values of the parameter $E^*$.

\section{Discussion}

\begin{figure}[ht]
\begin{center}
{
 \epsfig{file=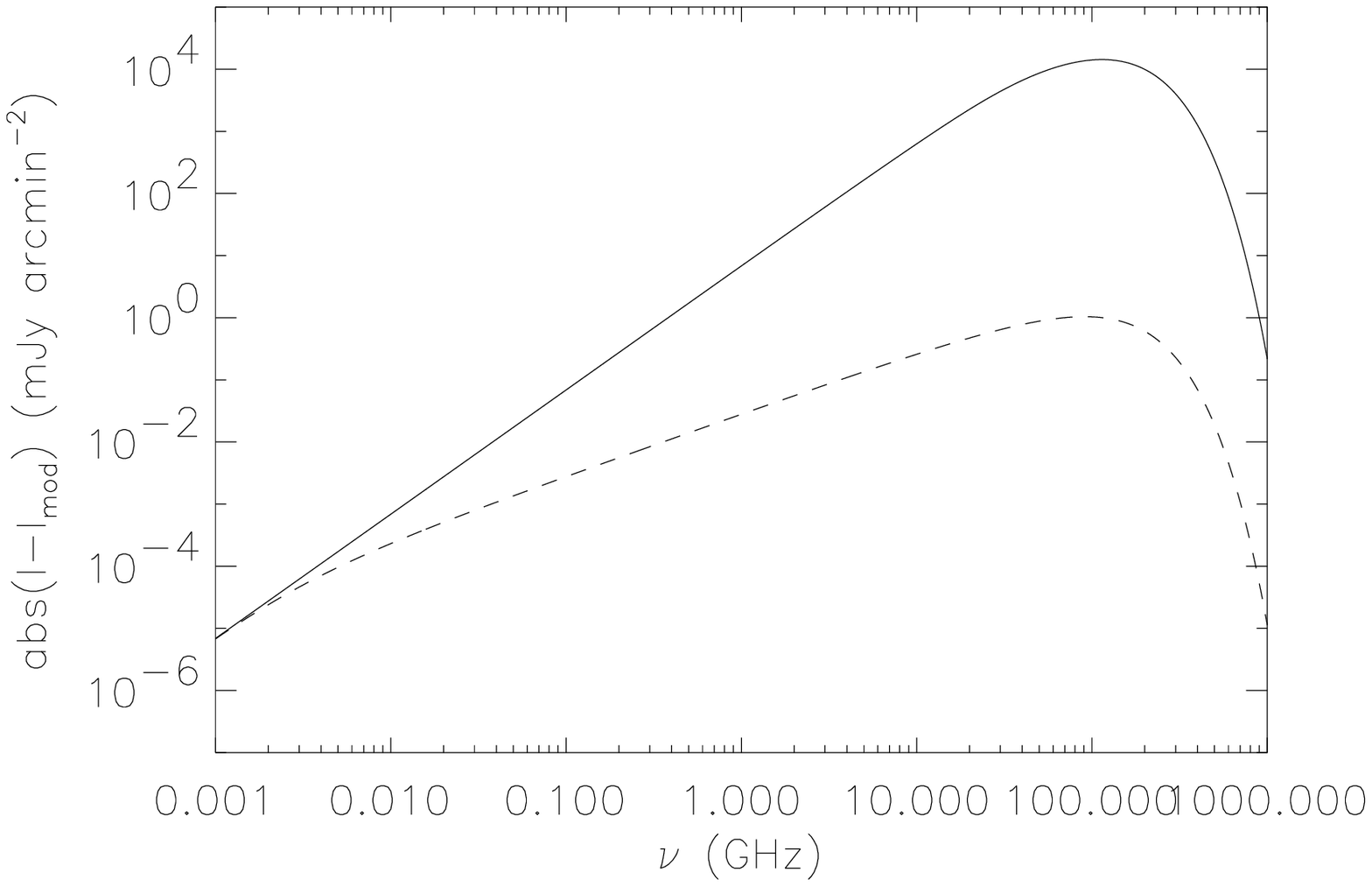,height=6.cm,width=9.cm,angle=0.0}
 \epsfig{file=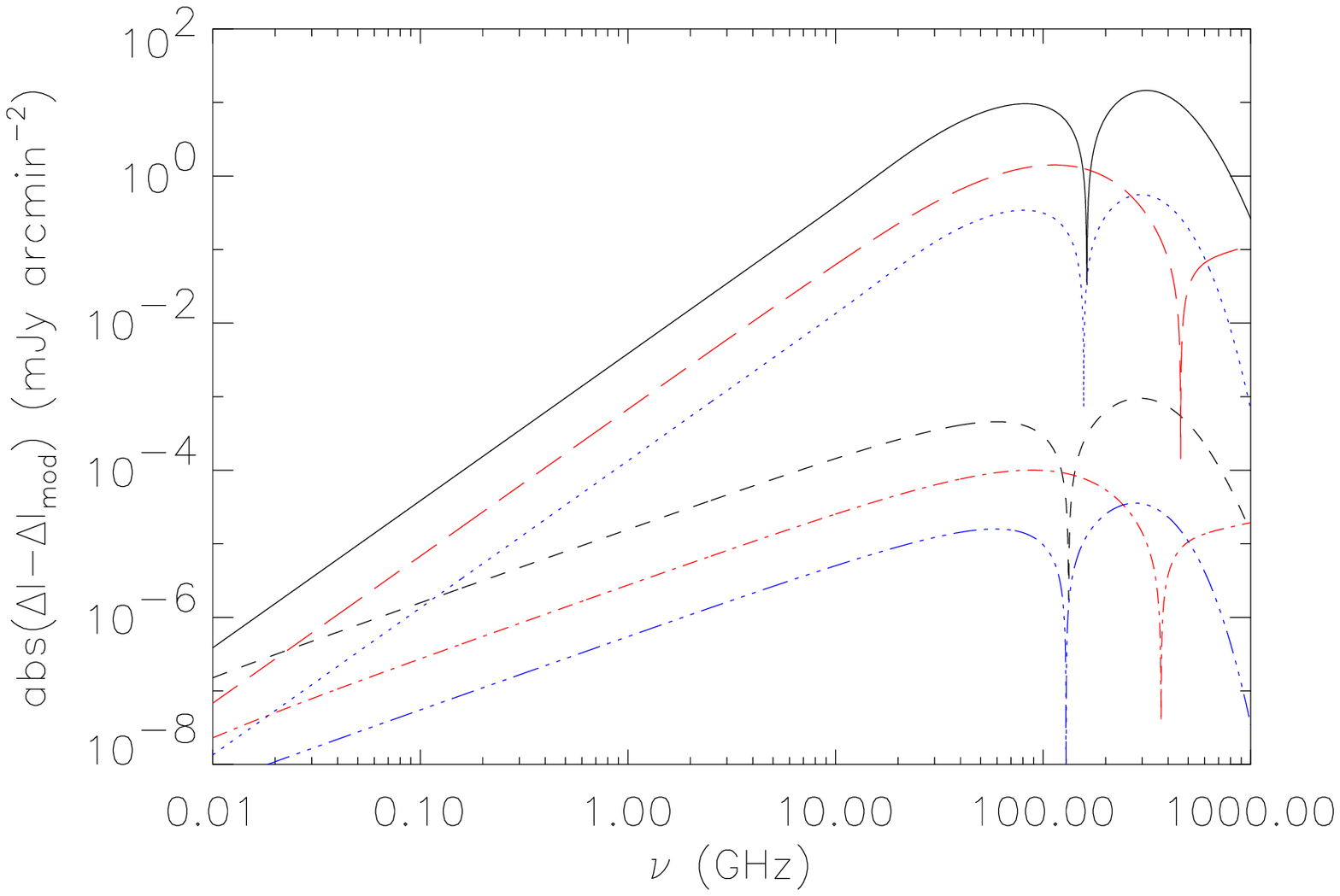,height=6.cm,width=9.cm,angle=0.0}
}
\end{center}
 \caption{\footnotesize{Upper panel: absolute difference between the standard CMB spectrum and the one modified for $E^*=10^{-4}$ eV (solid curve) and for $E^*=5\times10^{-9}$ eV (dashed curve).
Lower panel: absolute difference between the standard thermal SZE for $kT=15$ keV and $\tau=1\times10^{-2}$ and the one modified with 
$E^*=10^{-4}$ eV (solid curve) and  for $E^*=5\times10^{-9}$ eV (dashed curve),
between the standard non-thermal SZE (Colafrancesco et al. 2003) for $p_1=10$, $s=2.7$ and $\tau=1\times10^{-4}$ and the one modified with  $E^*=10^{-4}$ eV (red, long-dashed curve) and  for $E^*=5\times10^{-9}$ eV (red, dot-dashed curve),
and between the standard non-relativistic SZE for $y_0=1\times10^{-5}$ and the one modified with $E^*=10^{-4}$ eV (blue, dotted curve) and  for $E^*=5\times10^{-9}$ eV (blue, three dot-dashed curve).
 }}
 \label{err_abs}
\end{figure}

Figures \ref{szmod_15} and \ref{szmod_15_zoom} show that the distortion of the thermal SZE induced by the photon decay has a peculiar spectral characteristic. While at high frequencies ($\nu \simgt 220$ GHz) this distortion is similar to that of a non-thermal SZE (i.e. a shift of the null and the maximum of the SZE towards high frequencies, and hence difficult to disentangle; see, e.g.,  En\ss lin \& Kaiser 2000, Colafrancesco et al. 2003, Colafrancesco et al. 2009, Colafrancesco \& Marchegiani 2010, Colafrancesco et al. 2011), at low frequencies it has a quite unique spectral behavior, i.e. a shift of the minimum of the SZE towards higher frequencies and a very unique spectral shape at very low frequencies where the modified SZE can be positive in sign (we remind the reader that the standard SZE is always negative at all frequencies below the null, i.e. at $\nu \simlt 220$ GHz).
Therefore, observations of the SZE at very low frequencies (i.e., $\nu \simlt 50$ GHz) and/or around the minimum (i.e., in the range  100--150 GHz) can provide crucial information on the presence of a possible photon decay effect.\\ 
Figure \ref{err_abs} shows a comparison between the difference of the standard and modified CMB spectrum (for two different values of  $E^*$)  and the difference between the relative SZE spectra in a thermal and in a non-thermal case, as well as in the non-relativistic case.
For the CMB spectrum, the maximum difference is found around 100 GHz
(114 and 90.5 GHz for $E^*=10^{-4}$ and $5\times10^{-9}$ eV, where the difference between the CMB spectra is
$1.43\times10^4$ and 1.04 mJy arcmin$^{-2}$ respectively). 
For the thermal SZE the maximum difference is found in two distinct spectral regions: 
82.0 and 314 GHz for $E^*=10^{-4}$ eV, and 59.5 and 292 GHz for $E^*=5\times10^{-9}$ eV, where the differences between the SZE spectra
are 9.60 and 14.6 mJy arcmin$^{-2}$ for $E^*=10^{-4}$ eV, and $4.56\times10^{-4}$ and $9.50\times10^{-4}$ mJy arcmin$^{-2}$ for $E^*=5\times10^{-9}$ eV.
%
For the non-thermal SZE the frequencies at which the maximum difference if found are 
113 and $2.86\times10^4$ GHz for $E^*=10^{-4}$ eV
and 88.4 and $1.96\times10^4$ GHz for $E^*=5\times10^{-9}$ eV,
where the differences are 1.41 and 0.675 mJy arcmin$^{-2}$, 
and $1.00\times10^{-4}$ and $5.79\times10^{-5}$ mJy arcmin$^{-2}$, respectively.
In order to set more stringent constraints than the available ones on the photon decay it is necessary to have an instrumental sensitivity below these values in the two interesting frequency windows.
%

The CMB spectral distortion induced by the photon decay leads to a shift of the SZE minimum towards high frequencies. 
This effect is a unique signature of the photon decay because the position of the SZE minimum depends very weakly from any other additional astrophysical effects, like the non-thermal SZE (see, e.g., En\ss lin \& Kaiser 2000, Colafrancesco et al. 2003, Colafrancesco \& Marchegiani 2010).
%
%
We studied the frequency shift of the SZE minimum $\nu_{min}$ from its standard value 
of $\approx 128$ GHz with increasing $E^*$.
We found that this effect becomes evident only for $E^*>10^{-5}$ eV, 
that is much higher than the limit  $E^*<5\times10^{-9}$ eV derived from CMB spectral measurements (Heeck 2013):
the minimum frequency is 129, 131, 143 and 156 GHz for $E^*=10^{-6}$, $10^{-5}$, $5\times10^{-5}$ and 
$10^{-4}$ eV, respectively.

\begin{figure}[ht]
\begin{center}
{
 \epsfig{file=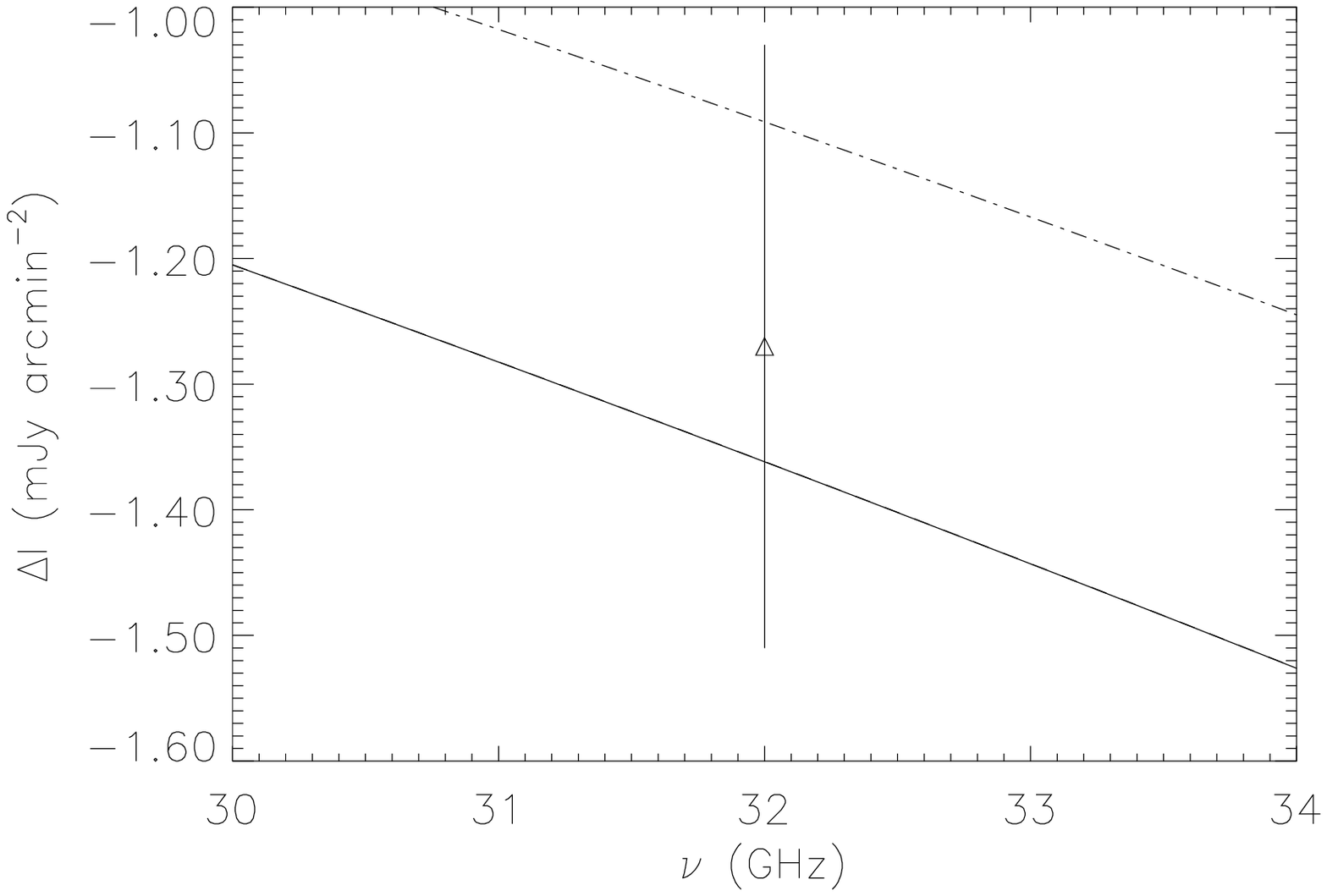,height=6.cm,width=9.cm,angle=0.0}
 \epsfig{file=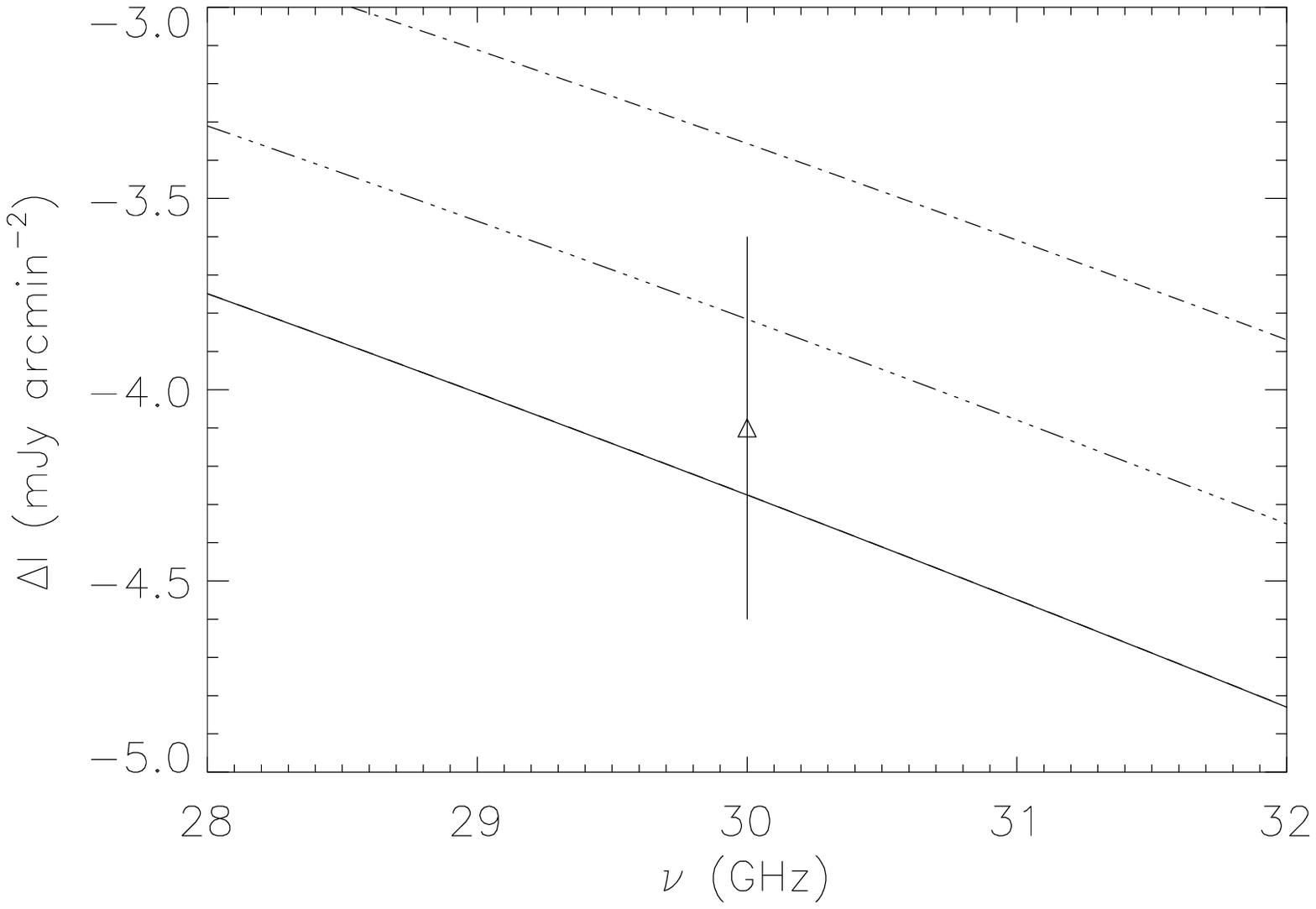,height=6.cm,width=9.cm,angle=0.0}
}
\end{center}
 \caption{\footnotesize{Upper panel: the SZE calculated for Coma (with
 $kT=8.2$ keV and $\tau=6.55\times10^{-3}$) for $E^*=0$ (solid),
 $5\times10^{-9}$ (dashed) and $1\times10^{-5}$ (dot-dashed) eV.
 Lower panel: the SZE calculated for A2163 (with 
 $kT=12.4$ keV and $\tau=1.56\times10^{-2}$) for $E^*=0$ (solid),
 $5\times10^{-9}$ (dashed), $5\times10^{-6}$ (3 dot-dashed)
 and $1\times10^{-5}$ (dot-dashed) eV. The Coma parameters are taken from 
 Hughes et al. (1989) and Briel et al. (1992), and from 
 Elbaz et al. (1995) and Markevitch et al. (1996) for A2163. 
 }}
 \label{szmod_ovro}
\end{figure}
\begin{figure}[ht]
\begin{center}
{
 \epsfig{file=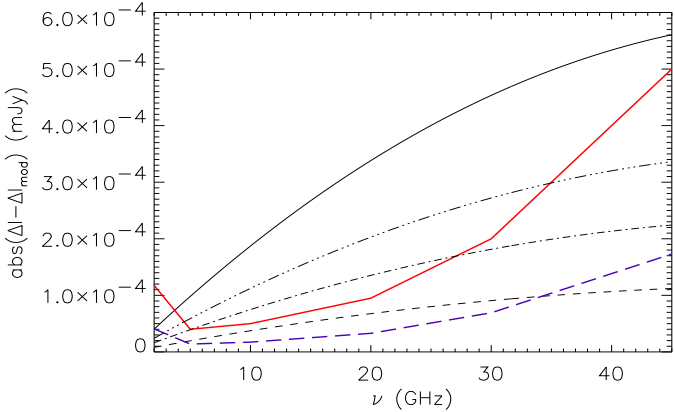,height=6.cm,width=9.cm,angle=0.0}
}
\end{center}
 \caption{\footnotesize{Difference between the standard  thermal SZE calculated for A2163 in 2--45 GHz range 
 and integrated in a 1 arcmin$^{2}$ area and the one modified for values $E^*=5\times10^{-9}$ eV (solid line), 
$3\times10^{-9}$ eV (dash - 3 dots), $2\times10^{-9}$ eV (dot-dashed) and $1\times10^{-9}$ eV (dashed).
The sensitivity achievable with SKA in $\sim30$ hours (red thick solid line)  and in $\sim260$ hours 
(blue thick long-dashed line) are shown (see Carilli 2005).
 }}
 \label{szmod_A2163_sens}
\end{figure}

Figure \ref{szmod_ovro} shows the comparison between the thermal SZE in the Coma and A2163 clusters calculated for various 
values of $E^*$ and the available SZE data from OVRO (for Coma; Herbig et al. 1995) and OVRO/Bima (for A2163 LaRoque et al. 2002). 
The OVRO data allow to set limit $E^*\simlt 7\times10^{-6}$ eV for A2163, and $E^*\simlt1.2\times10^{-5}$ eV 
for Coma. We conclude that the uncertainties of the OVRO data is much larger than the precision needed 
to set constraints on $E^*$ stronger or comparable to those of COBE-FIRAS on the CMB spectrum. 
%
Figure \ref{szmod_A2163_sens} shows the difference between the standard thermal SZE predicted for A2163 and the one modified by photon decay compared with the expected sensitivity of the Square Kilometer Array (SKA) for different values of  $E^*$.
We found that in 30 hours the SKA is able to measure the SZE difference at the 3.68, 2.21, 1.47 and 0.74$\sigma$ c.l.,
and in 260 hours it can measure a difference at the  10.6, 6.39, 4.26 and $2.13\sigma$ level c.l.  for $E^*=5$, 3, 2 and $1\times10^{-9}$ eV respectively,  or to put much more stringent upper limits on the value of $E^* \simlt 1.4\times10^{-9}$ eV
and $5\times10^{-10}$ eV for 30 and 260 hours respectively, and consequently set 
much more stringent constraints on the mass and decay time of the photon. 
The most effective spectral window for performing such measurements with the SKA is the one between 10 and 30 GHz. 
This frequency band is also advantageous because at these frequencies the synchrotron emission from radio halo diffuse emission 
in clusters (that has a quite steep spectrum, e.g. $\alpha_r \sim 1.2$ for A2163) is particularly low and hence not contaminating significantly the SZE measurement. 
The precise knowledge of the radio synchrotron spectrum at low and mid frequencies obtainable with the SKA will also allow to estimate its contribution at higher frequencies where the modified SZE is more evident. This contribution is anyway a conservative upper limit estimate because there is evidence that the radio halo spectrum steep ends at high frequencies (see, e.g., Thierbach et al. 2003 for the case of Coma): the subtraction of this synchrotron emission provides hence a conservative estimate (lower limit) of the distortion of the thermal SZE produced by the photon decay. 
%

Our analysis has been focussed on the effects of photon decay on the thermal SZE as computed in our general approach (Colafrancesco et al. 2003), leaving the discussion of the cluster peculiar motion effects or effects induced by line-of-sight variation of electrons properties to a further work. We note here that the corrections to the second order thermal SZE produced by the scattering induced anisotropies (see Chluba et al. 2013) can be at most $\sim 3-4$ times larger than the correction calculated in our approach. Because the second order correction to the SZE is of order $\tau^2$, even for a quite large value of $\tau\sim10^{-2}$ the relative change in the total SZE spectrum is at most of order $\sim 4\times10^{-4}$. In addition, we note that this correction becomes smaller for high electron temperatures (Chluba \& Dai 2013), and that the difference with our second order correction calculation is maximal around the minimum and the maximum regions of the SZE (Chluba et al. 2013).
Since we found that high temperature clusters, like A2163, and the low-frequency band $\nu \sim 10-50$ GHz provide the best cases to probe photon decay effect via the SZE, we can estimate that these last corrections do not affect appreciably our results. Finally, we stress that our main conclusions are obtained from the study of the difference between the original and the modified SZE, and because the scattering induced anisotropies corrections reflect on both the original and the modified spectra, the difference plotted in Fig.\ref{szmod_A2163_sens} is practically unaffected.

\section{Conclusions}

We have demonstrated in this paper that photon decay effects can be effectively studied by using the spectral distortions of the SZE in galaxy  clusters observable with the coming high sensitivity radio telescopes like SKA.
We have shown that measurements of the SZE in the range $\approx 10-50$ GHz are more competitive than CMB spectral measurements in order to set stringent constraints to the photon decay time, provided that we reach instrumental sensitivities of order of $\simlt 0.1 \mu$Jy. This frequency band is also the one less affected by other sources of astrophysical contamination and will be best explored with the advent of the high sensitivity SKA telescope. 
At higher frequencies (i.e., $\sim 120-180$ and $\sim 200-300$ GHz) there are other spectral windows where the SZE method is again competitive if not advantageous w.r.t. the CMB studies. The necessary sensitivity in the high frequency range can be achieved with the next coming Millimetron space mission (see, e.g., Colafrancesco 2012) even though other astrophysical sources of contamination (e.g. sources of non-thermal SZE and/or special distortions of the SZE due to multiple-temperature regions) could contaminate these measurements (see, e.g., Colafrancesco et al. 2011 and Prokhorov et al. 2012 for the case of the Bullet Cluster).
The complementarity between SKA and Millimetron SZE measurements will provide further and more stringent multi-frequency constraints on the effect of photon decay in the universe. The approved construction of both SKA and Millimetron will make these measurements realistic in the next few years.

\begin{acknowledgements}
S.C. acknowledges support by the South African Research Chairs Initiative of the Department of Science and Technology and National
Research Foundation and by the Square Kilometre Array (SKA). P.M. acknowledges support from the SKA post-graduate bursary initiative.
\end{acknowledgements}

{}

\end{document}